\RequirePackage{fix-cm}
\documentclass[smallcondensed]{svjour3}     
\smartqed  
\UseRawInputEncoding
\usepackage{graphicx}
\usepackage[dvipsnames]{xcolor}
\usepackage{amssymb}
\usepackage{float}
\usepackage{amsmath}
\usepackage[title]{appendix}
\newcommand{\newc}{\newcommand}
\newc{\beq}{\begin{equation}}
\newc{\eeq}{\end{equation}}
\newc{\kt}{\rangle}
\newc{\br}{\langle}
\newc{\beqa}{\begin{eqnarray}}
\newc{\eeqa}{\end{eqnarray}}
\newc{\pr}{\prime}
\newc{\longra}{\longrightarrow}
\newc{\ot}{\otimes}
\newc{\rarrow}{\rightarrow}
\newc{\h}{\hat}
\newc{\bom}{\boldmath}
\newc{\btd}{\bigtriangledown}
\newc{\al}{\alpha}
\newc{\be}{\beta}
\newc{\ld}{\lambda}
\newc{\sg}{\sigma}
\newc{\p}{\psi}
\newc{\eps}{\epsilon}
\newc{\om}{\omega}
\newc{\mb}{\mbox}
\newc{\tm}{\times}
\newc{\hu}{\hat{u}}
\newc{\hv}{\hat{v}}
\begin{document}

\title{One-Way Deficit and Holevo Quantity of Generalized $n$-qubit Werner State
}


\author{Yao-Kun Wang         \and
        Rui-Xin Chen  \and
        Li-Zhu Ge    \and
        Shao-Ming Fei  \and
        Zhi-Xi Wang
}


\institute{Yao-Kun Wang  \at
            College of Mathematics, Tonghua Normal University,  Tonghua, Jilin 134001, China\\
             \and
              Rui-Xin Chen  \at
           	  27 Jiefang Road, The High School Affiliated to Bohai University, Jinzhou, Liaoning 121000, China\\
           \and
            Li-Zhu Ge    \at
           	  The Branch Campus of Tonghua Normal University, Tonghua, Jilin 134001, China\\
           \and
           Shao-Ming Fei\at
             School of Mathematical Sciences,  Capital Normal University,  Beijing 100048,  China\\
           \and
           Zhi-Xi Wang \at
              \email{wangzhx@cnu.edu.cn}\\
              School of Mathematical Sciences,  Capital Normal University,  Beijing 100048,  China
}


\maketitle

\begin{abstract}
Originated from the work extraction in quantum systems coupled to a heat bath,
quantum deficit is a kind of significant quantum correlations like quantum entanglement. It links quantum thermodynamics with quantum information. We analytically calculate the one-way deficit of the generalized $n$-qubit Werner state. We find that the one-way deficit increases as the mixing probability $p$ increases for any $n$. For fixed $p$, we observe that the one-way deficit increases as $n$ increases. For any $n$, the maximum of one-way deficit is attained at $p=1$. Furthermore, for large $n$ ($2^n \rightarrow \infty$), we prove that the curve of one-way deficit versus $p$ approaches to a straight line with slope $1$. We also calculate the Holevo quantity for the generalized $n$-qubit Werner state, and show that it is zero.

\keywords{Generalized $n$-qubit Werner state \and one-way deficit \and Holevo quantity.}


\end{abstract}

\section{Introduction}
\label{intro}
Similar to quantum entanglement \cite{0} and quantum discord \cite{01,02}, quantum deficit \cite{1,2,3} is a kind of important nonclassical correlation, which characterizes the work extraction from a correlated system coupled to a heat bath by using nonlocal operations \cite{1}. Oppenheim et al. defined the work deficit \cite{1} to be the difference between the information of the whole system and the localizable information \cite{4}.
By means of relative entropy over all local von Neumann measurements on one subsystem, Streltsov et al. \cite{5,6} introduced the one-way deficit (OWD) as a resource of entanglement distribution. OWD is able to characterize quantum phase transitions in the XY model and even topological phase transitions in the extended Ising model \cite{40}. These results enlighten extensive researches of quantum phase transitions from the perspective of quantum information processing and quantum computation.
For a bipartite composite quantum system $\rho_{AB}$ associated with subsystems $A$ and $B$, the one-way deficit with respect to von Neumann measurement $\{\Pi_{k}\}$ on one subsystem is given by \cite{7}
\begin{eqnarray}
\Delta^{\rightarrow}(\rho_{AB})=\min\limits_{\{\Pi_{k}\}}
S(\sum\limits_{k}\Pi_{k}\rho_{AB}\Pi_{k})-S(\rho_{AB}),\label{definition}
\end{eqnarray}
where $S(\cdot)$ denotes the von Neumann entropy.

The Holevo bound characterizes the capacity of quantum states in classical communication \cite{8,9}. It is a keystone in many applications of quantum information theory\cite{nielsen,roga,lloyd,lupo,zhang,wu,guo}. With respect to the Holevo bound, the maximal Holevo quantity referred to weak measurements has been studied in \cite{wang}. The Holevo quantity of the SU(2)-invariant states has been investigated in \cite{wang2}.
The Holevo quantity of an ensemble $\{p_{i};\rho_{A|\Pi_i}\}$, corresponding to a bipartite quantum state $\rho_{AB}$ with the projective measurements $\{\Pi_i\}$ performed on the subsystem $B$, is given by \cite{wang}
\begin{eqnarray}\label{fo}
\chi\{\rho_{AB}|\{\Pi^B_i\}\}=\chi\{p_{i};\rho_{A|\Pi_i}\}\equiv S(\sum_{i}p_{i}\rho_{A|\Pi_i})-\sum_{i}p_{i}S(\rho_{A|\Pi_i}),
\end{eqnarray}
where
\begin{equation}\label{post}
p_i = \mbox{tr}_{AB}[(I_A \otimes \Pi_i ) \rho_{AB} ( {I}_A \otimes \Pi_i) ],~~
\rho_{A|\Pi_i} = \frac{1}{p_i} \mbox{tr}_B[({I}_A \otimes \Pi_i) \rho_{AB} ({I}_A \otimes \Pi_i)].
\end{equation}
It characterizes the A's accessible information about the B's measurement outcome when B projects the subsystem $B$ by the projection operators $\{\Pi^B_i\}$.

In this paper, we consider the generalized $n$-qubit Werner state given in \cite{siewert,cirac}. The state becomes the Werner state \cite{werner} when $n=2$. We study the OWD and the Holevo quantity under the bipartition of any single qubit ($B$ subsystem) and the remaining $n-1$ qubits ($A$ subsystem) for the generalized $n$-qubit Werner state. Here we perform a projective measurement on subsystem $B$. The general projective measurement operators are of the form,
\begin{align}\label{measure_op}
\Pi_1 = \mathbb{I}_A \otimes |u\kt_B {}_B\br u| && \text{and} && \Pi_2 = \mathbb{I}_A \otimes |v\kt_B {}_B\br v|,
\end{align}
where $|u\kt = \cos(\theta)|0\kt + e^{\textit{i}\phi}\sin(\theta)|1\kt$ and $|v\kt = \sin(\theta)|0\kt - e^{\textit{i}\phi}\cos(\theta)|1\kt$ with $0\leq\theta\leq\pi/2$ and $0\leq\phi\leq 2\pi$. In section \ref{owd}, we analytically calculate OWD between the subsystems $A$ and $B$, and derive the linear relationship between OWD and the mixing probability $p$ at the thermodynamic limit ($n\rightarrow \infty$). The Holevo quantity between the subsystems $A$ and $B$ is investigated in section \ref{negv}.

\section{OWD for generalized $n$-qubit Werner state}
\label{owd}
As the two-qubit Werner state is a special case of the Bell-diagonal states, while the quantum discord coincides with the one-way deficit for Bell-diagonal states \cite{wang1}, the one-way deficit is equal to the quantum discord for two-qubit Werner state \cite{01}. We first study the relation between the non-local correlations and the total number of qubits $n$. The generalized $n$-qubit Werner state is given as follows,
\begin{equation}
\rho_{W_{AB}} = p|\phi\kt_{AB}\hspace{0.03cm}{}_{AB}\br \phi| + \frac{(1-p)}{2^n}\mathbb{I}_{AB} , \label{n_qubit_werner1}
\end{equation}
where $0\leq p\leq 1$, $|\phi\rangle_{AB}$ is the $n$-qubit GHZ state under bipartition, $|\phi\rangle_{AB} = \frac{1}{\sqrt{2}} \left(\right.|0\kt^{\otimes n-1}_{A}|0\rangle_B + |1\kt^{\otimes n-1}_{A}$ $|1\rangle_B \left.\right)$, $\mathbb{I}_{AB}/2^n$ is the $n$-qubit maximally mixed state. To calculate the OWD between subsystems $A$ and $B$ for the state $\rho_{W_{AB}}$, we calculate the von Neumann entropy of $\rho_{W_{AB}}$.

Denote $N=2^n$. The matrix representation of the state $\rho_{W_{AB}}$ has the form,
\begin{equation}
\rho_{W_{AB}} = \begin{bmatrix}
a_{11}&0&0&\dots&a_{1 N}\\
0&a_{22}&0&\dots& 0\\
0&0&a_{33}&\dots& 0\\
\vdots& & &\ddots&\vdots\\
a_{N 1}& & & & a_{N N}
\end{bmatrix}_{N \times N}, \label{n_qubit_werner_matrix}
\end{equation}
where $a_{11} =a_{N N} =  \left(\frac{1-p}{2^n} + \frac{p}{2}\right)$, $a_{1 N} = a_{N 1} = \frac{p}{2}$ and $a_{22} = a_{33} = a_{44} = \ldots = a_{N-1\;N-1} = \left(\frac{1-p}{2^n}\right)$.
From the characteristic equation of the matrix $\rho_{W_{AB}}$,
$(a_{22}-\ld)(a_{33} - \ld)\cdots(a_{N-1\;N-1}-\ld)(\ld^2 - \ld(a_{11}+ a_{NN}) + a_{11}a_{NN} - a_{N1}a_{1N}) = 0$, we have the eigenvalues \cite{ram},
\begin{equation}
\ld_1 = \ld_2 = \ld_3= \ldots =\ld_{N-2} = \frac{1-p}{2^n}\label{eigv_N2}
\end{equation}
and
\begin{eqnarray}
\ld_{N-1} &=& \frac{1}{2}\left\{(a_{11}+a_{N N}) + \sqrt{(a_{11}-a_{N N})^2 + 4 a_{1 N}a_{N 1}}\right\}\nonumber \\
&=&\frac{1 + (2^n - 1)p}{2^n};\label{eigv_N1} \\
\ld_{N} &=& \frac{1}{2}\left\{(a_{11}+a_{N N}) - \sqrt{(a_{11}-a_{N N})^2 + 4 a_{1 N}a_{N 1}}\right\}\nonumber \\
&=&\frac{1-p}{2^n}.\label{eigv_N}
\end{eqnarray}%
Therefore, we have the entropy $S(\rho_{W_{AB}})$,
\begin{eqnarray}
S(\rho_{W_{AB}}) &=&  (N-2) \left\{-\left(\frac{1-p}{2^n}\right)\log_2 \left(\frac{1-p}{2^n}\right)\right\} -\left(\frac{1+(2^n -1)p}{2^n}\right) \nonumber \\ && \cdot \log_2 \left(\frac{1+(2^n -1)p}{2^n}\right) -  \left(\frac{1-p}{2^n}\right)\log_2 \left(\frac{1-p}{2^n}\right),\\
&=&- (2^n - 1) \left(\frac{1-p}{2^n}\right)\log_2 \left(\frac{1-p}{2^n}\right) -\left(\frac{1+(2^n -1)p}{2^n}\right) \nonumber \\ &&\cdot \log_2 \left(\frac{1+(2^n -1)p}{2^n}\right). \label{n_qubit_sab}
\end{eqnarray}

To compute $\min\limits_{\{\Pi_{k}\}}S(\sum\limits_{k}\Pi_{k}\rho_{W_{AB}}\Pi_{k})$ under measurement (\ref{measure_op}) on subsystem $B$, let us consider the following state,
\begin{eqnarray}\label{rho}
\rho&=&p\cos^2(\theta)|0\kt^{\otimes n-1}\br 0|^{\otimes n-1}  + pe^{\textit{i}\phi}\cos(\theta)\sin(\theta)|0\kt^{\otimes n-1}\br 1|^{\otimes n-1} \nonumber \\ && + pe^{-\textit{i}\phi}\cos(\theta)\sin(\theta)|1\kt^{\otimes n-1}\br 0|^{\otimes n-1} + p \sin^2(\theta)|1\kt^{\otimes n-1}\br 1|^{\otimes n-1}\nonumber \\ && + \sum_{i=0}^{2^{n-1}-1}\left(\frac{1-p}{2^{n-1}}\right)|i\kt\br i|. \label{n_qubit_ra1}
\end{eqnarray}

Denote $L=2^{n-1}$. The density matrix $\rho$ has the form,
\begin{equation}
\rho = \begin{bmatrix}
b_{11}&0&0&\dots&b_{1 L}\\
0&b_{22}&0&\dots& 0\\
0&0&b_{33}&\dots& 0\\
\vdots& & &\ddots&\vdots\\
b_{L 1}& 0 &0 &\dots & b_{LL}
\end{bmatrix}_{LL}, \label{n_qubit_ra1_matrix}
\end{equation}
where
\begin{align}
b_{11} &= p\cos^2{\theta} + \frac{1-p}{2^{n-1}}, & b_{L L} = p\sin^2{\theta} + \frac{1-p}{2^{n-1}},\\
b_{1 L} &= pe^{\textit{i}\phi}\cos(\theta)\sin(\theta), &
b_{L 1}= pe^{-\textit{i}\phi}\cos(\theta)\sin(\theta)
\end{align}
and
\begin{align}
b_{22} = b_{33} =\ldots= b_{L-1\;L-1} = \frac{1-p}{2^{n-1}}.
\end{align}
The eigenvalues of $\rho$ are
$\beta_1 = \beta_2 =\ldots=\beta_{L-2} = \frac{1-p}{2^{n-1}}$,
$\beta_{L-1} = \frac{1+(2^{n-1}-1)p}{2^{n-1}}$ and $\beta_L = \frac{1-p}{2^{n-1}}$.

From (\ref{measure_op}) and (\ref{rho}), it is direct to verify that $\Pi_1 \rho_{W_{AB}}\Pi_1=\frac{\rho}{2}\otimes |u\kt_B {}_B\br u|$ and $\Pi_2 \rho_{W_{AB}}\Pi_2=\frac{\rho}{2}\otimes |v\kt_B {}_B\br v|$.
Hence,
\begin{eqnarray}
\sum\limits_{k}\Pi_{k}\rho_{W_{AB}}\Pi_{k}&=&\Pi_1 \rho_{W_{AB}}\Pi_1+\Pi_2 \rho_{W_{AB}}\Pi_2\\
&=&\frac{\rho}{2}\otimes\left(|u\kt_B {}_B\br u|+|v\kt_B {}_B\br v|\right)\\
&=&\frac{\rho}{2}\otimes\left(
                          \begin{array}{cc}
                            1 & 0 \\
                            0 & 1 \\
                          \end{array}
                        \right).
\end{eqnarray}
From the calculation of the eigenvalues of $\rho$, we have the following eigenvalues of the matrix $\sum\limits_{k}\Pi_{k}\rho_{W_{AB}}\Pi_{k}$,
$\alpha_1 = \alpha_2 =\ldots=\alpha_{2L-4} = \frac{1-p}{2^{n}}$,
$\alpha_{2L-3}=\alpha_{2L-2} = \frac{1+(2^{n-1}-1)p}{2^{n}}$ and $\alpha_{2L-1}=\alpha_{2L} = \frac{1-p}{2^{n}}$.
The entropy of $\sum\limits_{k}\Pi_{k}\rho_{W_{AB}}\Pi_{k}$ is given by
\begin{eqnarray}
S(\sum\limits_{k}\Pi_{k}\rho_{W_{AB}}\Pi_{k}) &=&  - (2L - 4) \left(\frac{1-p}{2^{n}}\right)\log_2 \left(\frac{1-p}{2^{n}}\right) -2\left(\frac{1+(2^{n-1} -1)p}{2^{n}}\right) \nonumber \\ && \cdot \log_2 \left(\frac{1+(2^{n-1} -1)p}{2^{n}}\right) -  2\left(\frac{1-p}{2^{n}}\right)\log_2 \left(\frac{1-p}{2^{n}}\right)\nonumber\\
&=&- (2^{n-1} - 1) \left(\frac{1-p}{2^{n-1}}\right)\log_2 \left(\frac{1-p}{2^{n}}\right) -\left(\frac{1+(2^{n-1} -1)p}{2^{n-1}}\right) \nonumber \\ && \cdot \log_2 \left(\frac{1+(2^{n-1} -1)p}{2^{n}}\right). \label{min}
\end{eqnarray}

Note that $S(\sum\limits_{k}\Pi_{k}\rho_{W_{AB}}\Pi_{k})$ is independent of the parameters $\theta$ and $\phi$ in the measurement operators given in (\ref{measure_op}). Therefore, the minimization of $S(\sum\limits_{k}\Pi_{k}\rho_{W_{AB}}\Pi_{k})$ over the measurements is not required. Using Eqs. (\ref{definition}), (\ref{n_qubit_sab}) and (\ref{min}), we have the OWD of state $\rho_{W_{AB}}$,
\begin{eqnarray}
\Delta^{\rightarrow}(\rho_{W_{AB}})&=&
\min\limits_{\{\Pi_{k}\}}S(\sum\limits_{k}\Pi_{k}\rho_{W_{AB}}\Pi_{k})-S(\rho_{AB})\nonumber\\
&=&-(2^{n-1} - 1) \left(\frac{1-p}{2^{n-1}}\right)\log_2 \left(\frac{1-p}{2^{n}}\right) -\left(\frac{1+(2^{n-1} -1)p}{2^{n-1}}\right) \nonumber \\
& & \cdot \log_2 \left(\frac{1+(2^{n-1} -1)p}{2^{n}}\right)+(2^n - 1) \left(\frac{1-p}{2^n}\right)\log_2 \left(\frac{1-p}{2^n}\right) \nonumber \\
& &+\left(\frac{1+(2^n -1)p}{2^n}\right)\log_2 \left(\frac{1+(2^n -1)p}{2^n}\right).\label{deficit}
\end{eqnarray}

In Fig. 1, We plot the OWD as a function of $p$ for different number of qubits $n$. We observe that the OWD increases as $p$ increases for any $n$. As $n$ increases, the OWD increases for a given $p$, which indicates that the nonclassical correlations are dependent upon $n$. The maximum of OWD is attained at $p=1$ for any $n$.
\begin{figure}[t]
\includegraphics[scale=1.5]{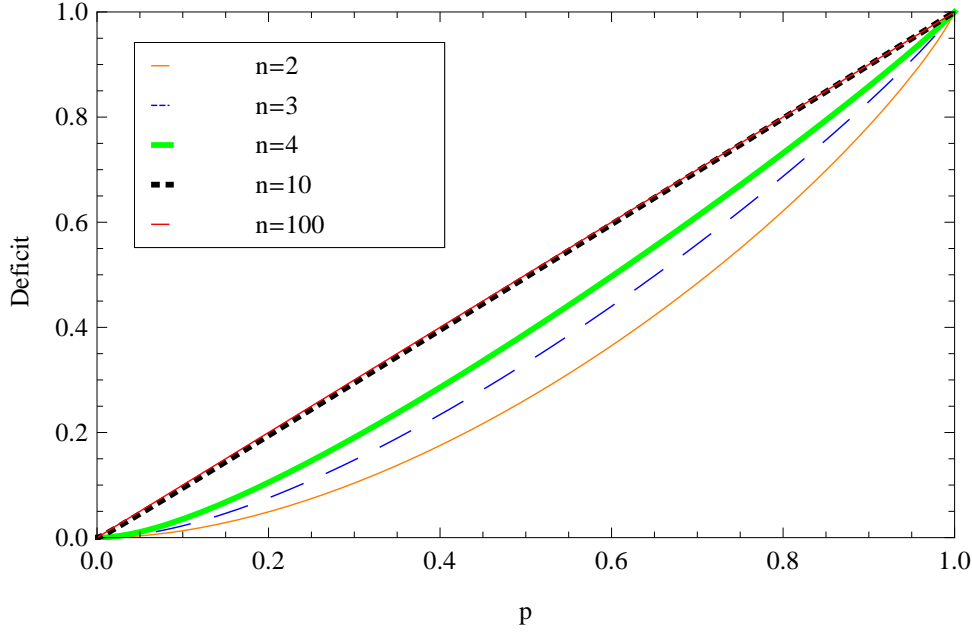}
\caption{(Color online) One-way deficit of the generalized $n$-qubit Werner State as a function of the mixing probability $p$ for different number of qubits $n$.}
\end{figure}

We next study the one-way deficit at thermodynamic limit ($n\rightarrow \infty$).
The OWD (\ref{deficit}) can be rewritten as,
\begin{eqnarray}
\Delta^{\rightarrow}(\rho_{W_{AB}})&=&-\left(1-p-\frac{1-p}{2^{n-1}}\right)\log_2 \left(\frac{1-p}{2^{n}}\right) -\left(\frac{1}{2^{n-1}}+p-\frac{p}{2^{n-1}}\right) \nonumber \\
& & \cdot \log_2 \left(\frac{1}{2^{n}}+\frac{p}{2}-\frac{p}{2^{n}}\right)+\left(1-p-\frac{1-p}{2^n}\right)\log_2 \left(\frac{1-p}{2^n}\right) \nonumber \\
& &+\left(\frac{1}{2^n}+p-\frac{p}{2^n}\right)\log_2 \left(\frac{1}{2^n}+p-\frac{p}{2^n}\right).
\end{eqnarray}
When $2^n \rightarrow\infty$, one obtains
\begin{eqnarray}
\lim_{2^n\to\infty}\Delta^{\rightarrow}(\rho_{W_{AB}})&=&-\left(1-p\right)\log_2 \left(\frac{1-p}{2^{n}}\right) -p\log_2 \left(\frac{p}{2}\right) \nonumber\\
& &+\left(1-p\right)\log_2 \left(\frac{1-p}{2^n}\right)+p\log_2 \left(p\right)\nonumber\\
&=& p. \nonumber
\end{eqnarray}
Interestingly, the change between OWD and $p$ saturates at thermodynamic limit. The curve in Fig. 1 approaches to a straight line with slope $1$, a phenomenon we call it ``saturation of one-way deficit".

\section{Holevo Quantity for generalized $n$-qubit Werner state}
\label{negv}
In this section, we calculate the Holevo quantity of the generalized $n$-qubit Werner state $\rho_{W_{AB}}$. Denote $p_i$ the probability with respect to the measurement outcome of $\Pi_i$, $i=1,2$. From (\ref{post}) we have
\begin{eqnarray}
p_1 &=& tr(\Pi_1\rho_{W_{AB}}\Pi_1)\nonumber\\
&=&tr(\frac{\rho}{2}\otimes |u\kt_B {}_B\br u|)=\frac{1}{2}. \nonumber
\label{n_qubit_p1}
\end{eqnarray}
The post measurement state of the subsystem $A$ is
\begin{eqnarray}
\rho_{A|\Pi_1} &=& \frac{1}{p_1}tr_B(\Pi_1\rho_{W_{AB}}\Pi_1)\nonumber\\
&=&\frac{1}{p_1}tr_B(\frac{\rho}{2}\otimes |u\kt_B {}_B\br u|)=\rho.\nonumber
\end{eqnarray}
Similarly, we have $p_2 = \frac{1}{2}$ and $\rho_{A|\Pi_2}=\rho$.

The Holevo quantity of the generalized $n$-qubit Werner state is then given by
\begin{eqnarray}\label{fo}
\chi\{\rho_{AB}|\{\Pi_i\}\}&=& S\left(\sum_{i}p_{i}\rho_{A|\Pi_k}\right)-\sum_{i}p_{i}S\left(\rho_{A|\Pi_k}\right)\nonumber\\
&=&S\left(\frac{1}{2}\rho+\frac{1}{2}\rho\right)-\left(\frac{1}{2}S(\rho)
+\frac{1}{2}S(\rho)\right)\nonumber\\
&=&0. \nonumber
\end{eqnarray}

\section{Conclusion} \label{con}
We have analytically calculated the one-way deficit of the generalized $n$-qubit Werner state, with the projective measurements performed on one-qubit subsystem. We have found that the OWD increases as $p$ increases for any $n$. When $n$ increases, the OWD increases for any fixed $p$. For any $n$, the maximum of OWD is attained at $p=1$. Furthermore, for large $n$ ($2^n \rightarrow \infty$), by analytical calculation we have proved that this curve OWD versus $p$ approaches to a straight line with slope $1$. We have also shown that the Holevo quantity of the generalized $n$-qubit Werner state is $0$.
\bigskip

\noindent{\bf Acknowledgments}\, \, This work is supported by the National Natural Science Foundation of China (NSFC) under Grant Nos. 12065021, 12075159 and 12171044; Beijing Natural Science Foundation (Grant No. Z190005); the Academician Innovation Platform of Hainan Province.

\bigskip
\footnotesize{\noindent\textbf{Conflict of Interest Statements} The authors declare no competing interests.}

\bigskip
\footnotesize{\noindent\textbf{Data Availability Statements} All data generated or analysed during this study are available from the corresponding author on reasonable request.}


%
\end{document}